\documentclass[preprint,amsfonts,amsmath,amssymb,eqsecnum,aps,showpacs,showkeys,floatfix]{revtex4}

\newcommand{\Gd}{\delta}

\newcommand{\Gg}{\gamma}

\newcommand{\Gl}{\lambda}

\newcommand{\GP}{\Phi}

\newcommand{\Gt}{\theta}

\newcommand{\Gx}{\xi}

\newcommand{\ew}{SU_W(2)\otimes U_{EM}(1)}
\newcommand{\newew}{U_{EW}(2)}
\newcommand{\oldew}{SU_I(2)\otimes U_Y(1)}
\newcommand{\sm}{SU_C(3)\otimes SU_W(2)\otimes U_{EM}(1)}
\newcommand{\D}{\not\!\!{D}}
\newcommand{\ket}[1]{\bigl|{#1}\bigr>}

\newcommand{\bs}[1]{\mathbf{#1}}
\newcommand{\Bold}[1]{\mbox{\boldmath$\mathit{#1}$}}
\newcommand{\sBold}[1]{\mbox{\boldmath$\mathit{\scriptstyle{#1}}$}}
\newcommand{\C}{\mathbb{C}}

\begin{document}

\title{\textbf{Quarks with Integer Electric Charge}}
\author{J. LaChapelle}



\begin{abstract}
Within the context of the Standard Model, quarks are placed in a
$(\mathbf{3},\mathbf{2})\oplus (\mathbf{3},\overline{\mathbf{2}})$
matter field representation of $\newew$. Although the quarks carry
unit intrinsic electric charge in this construction, anomaly
cancellation constrains the Lagrangian in such a way that the
quarks' associated currents couple to the photon with the usual
$2/3$ and $1/3$ fractional electric charge associated with
conventional quarks. The resulting model is identical to the
Standard Model in the $SU_C(3)$ sector: However, in the $\newew$
sector it is similar but not necessarily equivalent. Off hand, the
model appears to be phenomenologically equivalent to the
conventional quark model in the electroweak sector for
experimental conditions that preclude observation of individual
constituent currents. On the other hand, it is conceivable that
detailed analyses for electroweak reactions may reveal
discrepancies with the Standard Model in high energy and/or large
momentum transfer reactions.

The possibility of quarks with integer electric charge strongly
suggests the notion that leptons and quarks are merely different
manifestations of the same underlying field. A speculative model
is proposed in which a phase transition is assumed to occur
between $SU_C(3)\otimes U_{EM}(1)$ and $U_{EM}(1)$ regimes. This
immediately explains the equality of lepton/quark generations and
lepton/hadron electric charge, relates neutrino oscillations to
quark flavor mixing, reduces the free parameters of the Standard
Model, and renders the issue of quark confinement moot.

\end{abstract}



\maketitle

\section{Introduction}
The basis of this paper is the realization that the intrinsic
charges carried by a matter field and the coupling strengths of
its associated currents are not necessarily equivalent. This
realization was expounded in \cite{LAC1} where the quantum
numbers characterizing elementary fields and gauge/matter
field coupling strengths were analyzed for gauge theories with
direct product groups.

As an illustrative example, consider a theory with internal
symmetry $SU(3)\otimes SU(2)\otimes U(1)$ and suppose that the
matter field $\mathit{\Psi}$ characterized by the quantum numbers
$(C,I,Y)$ and its $SU(2)\otimes U(1)$ conjugate field
$\mathit{\Psi}'$ characterized by the quantum numbers
$(C,\overline{I},\overline{Y})=(C,I,-Y)$ are elementary fields
furnishing the representation
$(\mathbf{3},\mathbf{2},\mathbf{1})\oplus
(\mathbf{3},\overline{\mathbf{2}},\overline{\mathbf{1}})$. The
Lagrangian density will contain two terms of the form
$\alpha^2(\overline{\mathit{\Psi}}\D \mathit{\Psi})$ and
$\beta^2(\overline{\mathit{\Psi}'}\D' \mathit{\Psi}')$ with
$\alpha ,\beta\in\mathbb{R}$ and $\alpha^2+\beta^2=1$. In the
absence of the $SU(3)$ symmetry, $\alpha^2$ and $\beta^2$ are
necessarily equal and can be absorbed into a redefinition of the
matter fields. However, in some circumstances $\alpha^2\neq
\beta^2$ and a renormalization of the matter fields cannot absorb
the relative factor of $\alpha/\beta$ since $\mathit{\Psi}$ and
$\mathit{\Psi}'$ are $SU(2)\otimes U(1)$ conjugate.

It is clear that $\mathit{\Psi}$ carries the $U(1)$ charge $Y$ yet
its associated current couples to the $U(1)$ gauge field with
strength $\alpha^2 Y$. Likewise, $\mathit{\Psi}'$ carries charge
$-Y$ but its associated current couples with strength $-\beta^2
Y$. Note, however, that the $SU(3)$ `charges' and coupling
strengths are equivalent because $\alpha^2+\beta^2=1$. Also note
that this phenomenon does not occur if only the
$(\mathbf{3},\mathbf{2},\mathbf{1})$ matter field is included.

To see how this can be applied to a specific model of hadronic
constituents (HC), it is best to first recall the historical
motivation leading to the conventional assignment of quark quantum
numbers.

The Standard Model (SM) began as an electroweak theory of leptons
\cite{G,W1,S}. Later, hadrons where incorporated by considering
the known structure of the charged hadronic current, the
postulated quark composition of hadrons, and the assumed isospin
and hypercharge quark quantum numbers \cite{GIM,FGL,W2}.

The canonical status enjoyed by the isospin and hypercharge
quantum numbers of quarks can be attributed to the structure of
the $\oldew$ symmetry (sub)group and the success of the
Gell-Mann/Nishijima relation ($Q\propto T+1/2Y$) in classifying
mesons and baryons in various approximate isospin and flavor
symmetry models. Historically, this led to the conclusion that the
$(u,d,s)$ quarks possessed fractional electric charge. Including
the $SU_C(3)$ symmetry in the SM, assuming fractional electric
charge, and using the Gell-Mann/Nishijima relation leads naturally
to the conventional assignment of quarks in the
$(\mathbf{3},\mathbf{2},\mathbf{1/3})$ representation of
$SU_C(3)\otimes\oldew$.

Now, suspending momentarily our notion of quarks and their assumed
isospin and hypercharge, imagine HC\footnote{The term
hadronic constituents(HC) is used in order to keep a clear
distinction from the standard quarks.} corresponding to a matter
field representation in the unbroken electroweak symmetry domain
of the SM, viz. $\ew$. A natural assignment for the HC is a
$(\mathbf{2},\mathbf{1})$ field and a
$(\overline{\mathbf{2}},\overline{\mathbf{1}})=(\overline{\mathbf{2}},-\mathbf{1})$
anti-field. (By natural we mean that there exists a preferred basis
in the Lie algebra in which the charged gauge bosons have an
electric charge of $\pm e$; and one might expect the gauge bosons
exchange this electric charge quanta with elementary matter
fields.) Now assign the HC to an $SU_C(3)$ triplet without
recourse to the Gell-Mann/Nishijima relation. Should the HC matter
fields furnish a $(\mathbf{3},\mathbf{2},\mathbf{1})$, a
$(\mathbf{3},\overline{\mathbf{2}},-\mathbf{1})$, or a combination of the
two? It is not unreasonable to expect a combination.

An apparent contradiction arises immediately: how can a color
triplet of HC, which possess integer electric charge, combine to
form hadrons with their observed electric charges? The answer is
that the intrinsic electric charge carried by an elementary matter
field and its associated coupling strength to a gauge boson are
not necessarily equivalent by the mechanism explained above if
both $(\mathbf{3},\mathbf{2},\mathbf{1})$ and
$(\mathbf{3},\overline{\mathbf{2}},-\mathbf{1})$ HC fields are included.

It turns out that an appropriate combination of
$(\mathbf{3},\mathbf{2},\mathbf{1})$ and $
(\mathbf{3},\overline{\mathbf{2}},-\mathbf{1})$ HC matter terms can be
implemented within the context of the SM Lagrangian, and anomaly
cancellation uniquely determines the relative factors $\alpha^2$
and $\beta^2$ in the terms. (Henceforth, the HC
based on this new representation will be called `iquarks' in order to clearly
differentiate between the new quarks with integer electric charge
and conventional quarks.) Consequently, the iquark matter fields
couple to the electroweak gauge bosons with fractional coupling
strengths reminiscent of conventional quark couplings.
Specifically, the electromagnetic current contains the expected
$2/3e$ and $-1/3e$ factors even though the iquarks have integer
intrinsic electric charge.

With this matter field representation, the usual predictions of
the SM that do not depend on iquark electroweak currents, as well
as anomaly cancellation and resolution of the $\pi^{\mathrm
0}\rightarrow{2\Gg}$ problem, are exactly maintained. The iquark
electroweak currents can be cast in terms of conventional quarks
by identifying the up and down quark with an iquark doublet and
its electroweak anti-doublet within each generation. It follows
that conventional quarks are, in a sense, an average of two
iquarks. Insofar as experiments are not able to distinguish
individual iquark currents, it appears that the predictions of
this model for the electroweak sector will coincide with those of
the SM in terms of conventional quarks.

Despite superficial appearances, there is a possibility that the
iquark electroweak sector could yield results that differ from the
SM --- especially at high energy where individual currents might be
distinguished due to mass differences of the iquarks. However,
without detailed analyses of reaction rates (which is not
undertaken here), the question of whether electroweak predictions
of this model differ from the SM predictions remains open.

The choice of iquark representation advocated here is interesting
since it: (i) reproduces many, if not all, of the successful
predictions of the SM, (ii) may lead to experimentally verifiable
differences, and (iii) suggests a closer kinship between hadronic
constituents and leptons. In fact, the close kinship leads to the
conjecture that leptons and iquarks are different manifestations
of the same matter field. The idea is that for certain regions of
field phase space, ostensibly characterized by configurations with
particle content depending on space-time separation and relative
four-momentum, there is a phase change.

A relationship between leptons and iquarks would achieve an
economy of elementary particles and free parameters as well as
suggest new models for extensions of the SM. It is interesting to
note that neutrino oscillations would imply flavor mixing in the
iquark electroweak currents which leads to the hope of gleaning
some relationship between matter field masses, QCD and the
Kobayashi-Maskawa matrix.

It should be mentioned that quarks with integer charge have been
proposed before (see, e.g., \cite{GIM}, \cite{PS}, and the review
of \cite{GS}). However, the symmetry groups of these models are
not the $\sm$ of the SM,  and the iquark model presented here is
neither related to these models nor inspired by them. Also, the
proposed iquarks are not ``preons'' or``pre-quarks'' (see, e.g.,
\cite{T1,PS2,AH} and the review of \cite{T2}). That is,
conventional quarks are not composite states of the iquarks.
Instead, within this framework, conventional quarks can be
interpreted as an \emph{average} description of the iquarks.

\section{Intrinsic Charge and Coupling Strength}\label{charge and coupling
strength}

Before considering the specific model, it is helpful to examine
the relationship between intrinsic charges and coupling strengths
associated with an internal symmetry group.\footnote{This section
is a condensed version of the general analysis of \cite{LAC1}.}
The special case under consideration is a gauge field theory with
direct product group $K=G\times H$ where $G$ and $H$ are simple
compact and/or $U(1)$ Lie groups. Let the associated Lie algebras,
$\mathcal{G}$ and $\mathcal{H}$ respectively, be generated by the
bases $\{\bs{g}_i\}_{i=1}^{\mathrm{dim}G}$ and
$\{\bs{h}_r\}_{r=1}^{\mathrm{dim}H}$.

Suppose there exist matter fields that furnish inequivalent
irreducible representations (irreps) $\rho^{(a)}(G)$ and
$\rho^{(b)}(H)$. Then the inequivalent irreps of the direct
product group $\rho^{(a\times b)}(GH)=\rho^{(a)}(G)\otimes
\rho^{(b)}(H)$ are comprised of \emph{all} combinations of $a$ and
$b$. The associated representations of the Lie algebra
$\mathcal{K}=\mathcal{G}\oplus\mathcal{H}$ are ${\rho_e^{(a\times
b)}}'(\bs{g}_i+\bs{h}_r)={\rho_e^{(a)}}'(\bs{g}_i)
\oplus{\rho_e^{(b)}}'(\bs{h}_r)$ where $\rho_e'$ is the derivative
map of the representation evaluated at the identity element. It is
$\rho_e'(\mathcal{K})$ that determines the normalization of the
gauge fields via an inner product and the gauge/matter field
interactions via the covariant derivative $\D$.

Now, the scale of the intrinsic charges carried by the gauge
fields is determined by an inner product on the Lie algebra. The
scale ambiguity of the inner product for each simple compact and
$U(1)$ subgroup contributes an adjustable parameter that can be
absorbed into the definition of the gauge field. The interaction
of the gauge fields with the matter fields, as encoded in the
(renormalized) covariant derivative, results in an exchange of
these charge quanta; and this characterizes the intrinsic charges
of the matter fields.

On the other hand, the coupling strengths between the gauge fields
and matter currents are determined by the specific form of the
matter field Lagrangian. The most general spinor matter field
Lagrangian density consistent with the requisite symmetries
consists of a sum over the inequivalent irreps of the direct
product group:
\begin{equation}
  \mathcal{L}_m=i\sum_{a,b}\kappa_{ab}\overline{\psi}^{(a\times b)}
  \cdot\D\psi^{(a\times b)}+\mbox{mass terms}
\end{equation}
where $\kappa_{ab}$ are positive real constants that are
constrained by various consistency conditions.

These matter field terms give rise to the covariantly conserved
matter field currents
\begin{equation}
  j_{(i,r)}^{\mu}=\sum_{a,b}\kappa_{ab}\overline{\psi}^{(a\times b)}
  \cdot\gamma^{\mu}{\rho_e^{(a\times
b)}}'(\bs{g}_i+\bs{h}_r)\psi^{(a\times b)}\;.
\end{equation}
Evidently, the ratios of coupling strengths and associated
intrinsic charges are given by the $\kappa_{ab}$. It is clear that
$\kappa_{ab}$ can be absorbed by a field redefinition if either:
(i) $G=I$ or $H=I$, or (ii) the matter fields are not related
somehow. Otherwise, non-trivial $\kappa_{ab}$ may persist.

\section{Hadronic Constituents} \label{hadronic constituents}
Since the local details of the SM depend on the Lie algebra of the
gauge group --- insofar as the Lagrangian density is concerned --- and
since $su(2)\oplus u(1)\cong u(2)$, we may as well use
$U(2)=SU(2)\otimes U(1)/Z_2$ instead of $SU(2)\otimes U(1)$ for
the electroweak gauge group. Moreover, the emphasis on electric
charge suggests using $\newew$, which is characterized by the Lie
algebra decomposition $su_W(2)\oplus u_{EM}(1)$.
 The idea now is to have the iquark
matter fields furnish the same $\newew$ representation as the
lepton matter fields. Since the iquarks are to have integer
charge, there must be some mechanism to effect fractional
couplings in the electroweak currents. The solution is to consider
a combination of a $(\mathbf{3},\mathbf2)$ \emph{and} a
$(\mathbf{3},\overline{\mathbf2})$ of $SU_C(3)\otimes
U_{EW}(2)$\footnote{This combination was initially proposed in
\cite{LAC2}.}. Anomaly cancellation determines the allowed
combination, and the necessary fractional couplings ensue.
\vspace{.25in}

\noindent\textbf{Remark}: There are good reasons to believe the
electroweak group is $U(2)$. First, if $\rho$ is a representation
of $SU(2)\otimes U(1)$ furnished by the lepton fields of the SM,
then $\mathrm{ker}\rho=Z_2$ and therefore the lepton matter fields
do not furnish a faithful representation\cite{I}. The group
that does act effectively on the matter fields is $SU(2)\otimes
U(1) /Z_2=U(2)$. (Recall that we require faithful
representations.) Second, both $SU(2)\otimes U(1)$ and $U(2)$ have
the same covering group $Gl(1,q)$. Representations of $Gl(1,q)$
will descend to representations of $SU(2)\otimes U(1)$ or $U(2)$
if the associated discrete factor groups are represented trivially, i.e., by the unit matrix. For $SU(2)\otimes U(1)$ this
requirement implies no relationship between isospin and
hypercharge, but for $U(2)$ it implies $n=T+1/2Y$ with $n$ integer\cite{GIL}. Identifying $n$ with electric charge renders the
Gell-Mann/Nishijima relation and electric charge quantization a
consequence of the group $U(2)$. Third, symmetry reduction from
$U(2)$ to $U(1)$ is less constrained than reduction from
$SU(2)\otimes U(1)$ \cite{I}. Fourth, from a fiber bundle point
of view, the most general structure group for a matter field
doublet defined on a paracompact base space is $U(2)$.

\subsection{The Model}
First and foremost, we require the iquark matter fields to furnish
inequivalent faithful irreps of $SU_C(3)\times\newew$ for physical
reasons and so that the results of \cite{LAC1} can be applied.

For the iquark matter field sector of the model, the general
(classical) setup begins with a principal bundle with structure
group $SU_C(3)\times\newew$ along with associated vector bundles
over Minkowski space-time $V_{\sBold{R}}\rightarrow M^4$ where
$\Bold{R}$ designates the representation furnished by a particular
matter field. The typlical fiber of $V_{\sBold{R}}$ will depend on
$\Bold{R}$. For example, for the
$(\mathbf{3},\mathbf2)\oplus(\mathbf{3},\overline{\mathbf2})$ of
$SU_C(3)\times\newew$ the typical fiber is
$\mathbb{C}^{\,\text{3}}\otimes(\mathbb{C}^{\,\text{2}}
\oplus\mathbb{C}^{\,\text{2}})$. Here $\mathbb{C}^{\,\text{3}}$
carries the fundamental representation of $SU_C(3)$, and
$\mathbb{C}^{\,\text{2}}\oplus\mathbb{C}^{\,\text{2}}$ carries the
$\newew$ fundamental representation and its conjugate
representation. The internal degrees of freedom of elementary
matter fields are (by definition) identified with a chosen basis
of the typical fiber of $V_{\sBold{R}}$. Local gauge symmetry
allows for a consistent choice of basis at each space-time point.

Spinor matter fields require the product bundle $S\otimes
V_{\sBold{R}}$ where $S$ is a spinor bundle over Minkowski
space-time. For example, given a trivialization of $S\otimes
V_{(\mathbf{3},\mathbf2)}$, let $\{\Bold{e}_{\alpha
Aa}\}:=\{\Bold{\psi}_{\alpha}\otimes\Bold{e}_A\otimes\Bold{e}_a\}$
be the chosen basis that spans the typical fiber
$\C^4\otimes\C^{\,\text{3}}\otimes\C^{\,\text{2}}$. (Indices are
assumed to have the necessary ranges for any given
representation.) Sections $\Bold{\Psi}=\mathit{\Psi}^{\alpha
Aa}\Bold{e}_{\alpha Aa}$ of $S\otimes V_{(\mathbf{3},\mathbf2)}$
constitute the elementary spinor fields in the
$(\mathbf{3},\mathbf2)$ representation, and
$\Bold{e}_A\otimes\Bold{e}_a$ encode the internal
$SU_C(3)\times\newew$ degrees of freedom. For the conjugate
representation $(\mathbf{3},\overline{\mathbf2})$, we have
$\widetilde{\Bold{\Psi}}=\mathit{\Psi}^{\alpha
A\bar{a}}\Bold{e}_{\alpha
A\bar{a}}:=[i\tau_2]^{\bar{a}}_a\mathit{\Psi}^{\alpha
Aa}(\Bold{\psi}_{\alpha}\otimes\Bold{e}_A\otimes\Bold{e}_a^*)$.
There are analogous expressions for elementary fields furnishing
the other representations.

The first step is to assign representations to the iquark matter
fields. In order to simplify the presentation, Lorentz/spinor
components of the fields will be suppressed since they are just
treated in the usual manner. So attention will be restricted to
the sub-bundles $V_{(\mathbf{3},\mathbf2)}$ and
$V_{(\mathbf{3},\mathbf1)}$ over $M^{4}$ with structure group
$SU_C(3)\times\newew$.

Analogy with lepton matter fields suggests defining the
$(\mathbf{3},\mathbf2)$ iquark $\Bold{H}^{+}:=H_+^A\Bold{e}_A$ and
its $\newew$ conjugate $(\mathbf{3},\overline{\mathbf2})$
$\Bold{H}^-:={H_-^A}{\Bold{e}_A}$ by
\begin{eqnarray} \label{matter fields}
H_{+}^{A}:=\mathit{\Psi}^{Aa}\Bold{e}_a
=\mathit{\Psi}^{A1}\Bold{e}_1+\mathit{\Psi}^{A2}\Bold{e}_2
=:(h^+\Bold{e}_1)^A+ (\xi^0\Bold{e}_2)^{A}
=\begin{pmatrix} h^{+}\\
\xi^{0}
\end{pmatrix}^{A}
\end{eqnarray}
and \begin{eqnarray}\label{matter fields 2}
H_{-}^{A}:=[i\tau_2]^a_b\mathit{\Psi}^{Ab}\Bold{e}^*_a
=\mathit{\Psi}^{A2}\Bold{e}^*_1-\mathit{\Psi}^{A1}\Bold{e}^*_2
={\mathit{\Psi}^{A2}}^\ast\Bold{e}_1-{\mathit{\Psi}^{A1}}^*\Bold{e}_2
=:\begin{pmatrix} {\xi^{0}}^\ast\\ h^{-}
\end{pmatrix}^{A}\;.
\end{eqnarray}
Here $h^{\pm}$ and $\Gx^0$ are complex space-time Dirac spinor
fields (the superscripts denote electric charge) and
$\Bold{e}_{1,2}$ span $\C^2$.\footnote{Since there is no electroweak charge to distinguish $\Gx^0$ from ${\Gx^0}^\ast$ we will not include the $\ast$-designation from here on.} The $h^{\pm}$ and $\xi^0$ comprise
the iquarks.\footnote{The asignment
$h^-\Bold{e}_{\bar{2}}:=-h^+\Bold{e}^*_2$ follows from the
representation $\Bold{Q}^{-}(h^{-}\Bold{e}_{\bar{2}})=-
h^{-}\Bold{e}_{\bar{2}}$ of the electric charge generator on
$V_{(\mathbf{3,\overline{2}})}$ (see eq. (\ref{representation})).}
There are three copies of $\Bold{H}^{\pm}$ accounting
for the three iquark generations. No generality is sacrificed by
assuming $\Bold{H}^{\pm}$ are normalized.

By assumption, both the left and right-handed iquarks furnish the
$\mathbf3$ of $SU_C(3)$. The left-handed iquarks furnish the
$\mathbf2$ and $\overline{\mathbf2}$ and the right-handed iquarks
the $\mathbf1^+$, $\mathbf1^-$ and $\mathbf1^0$ of $\newew$. Thus,
we have $\Bold{H}^+_\mathrm{L}:=(\Bold{H}^+)_\mathrm{L}$,
$\Bold{H}^-_\mathrm{L}$, $\Bold{h}^+_\mathrm{R}$,
$\Bold{h}^-_\mathrm{R}$, and $\Bold{\xi}^0_\mathrm{R}$ furnishing
the $(\mathbf{3},\mathbf2)$, $(\mathbf{3},\overline{\mathbf2})$,
$(\mathbf{3},\mathbf1^+)$, $(\mathbf{3},\mathbf1^-)$, and
$(\mathbf{3},\mathbf1^0)$ respectively.

Next, the gauge potential must be specified in the relevant
representations. In the broken $\newew$ symmetry regime, which is
characterized by matter fields with conserved electric charge, the
Lie algebra $u_{EW}(2)$ decomposes as $u_{EM}(1)\oplus
(u_{EW}(2)/u_{EM}(1))$. Thus the gauge bosons are also
characterized by electric charge. This implies that the broken
symmetry generators are eigenvectors of the adjoint map of the
unbroken, electric charge generator. That is, the Lie algebra
decomposition is $u_{EW}(2)=u_{EM}(1)\oplus \mathrm{k}$ such that
$u_{EM}(1)\cap\mathrm{k}=0$ and
$ad(u_{EM}(1))\mathrm{k}\subseteq\mathrm{k}$. Since $\newew$ has
rank $2$, the relevant basis is
\begin{eqnarray}\label{basis}
&[\bs{e}_{\pm},\bs{e}_{\mp}]&=\sum_{i}\pm {c'}_i\bs{h}_i,\nonumber\\
&[\bs{e}_{\pm},\bs{h}_i]&=\pm c_i\bs{e}_{\pm},\nonumber\\
&[\bs{h}_{i},\bs{h}_j]&=0,
\end{eqnarray}
where $\{\bs{e}_+,\bs{e}_-,\bs{h}_1,\bs{h}_2\}$ spans $u_{EW}(2)$,
and $c_i,{c'}_i$ are constants with $i,j\in\{1,2\}$. The most
general $2$-dimensional representation allowed by (\ref{basis}) is
generated by
\begin{eqnarray}\label{rep}
  \bs{T}_+:=\rho'_e(\bs{e}_+)&=&
  i\begin{pmatrix}
  0 & & t \\
  0 & & 0
  \end{pmatrix},\nonumber\\
  \bs{T}_-:=\rho'_e(\bs{e}_-)&=&i\begin{pmatrix}
  0 & & 0 \\
  t & & 0
  \end{pmatrix},\nonumber\\
  \bs{T}_0:=\rho'_e(\bs{h}_1)&=&i\begin{pmatrix}
  u & & 0 \\
  0 & & v
  \end{pmatrix},\nonumber\\
  \bs{Q}:=\rho'_e(\bs{h}_2)&=&i\begin{pmatrix}
  r & & 0 \\
  0 & & s
  \end{pmatrix},
\end{eqnarray}
where $r,s,t,u,v$ are real constants, $\rho:\newew\rightarrow
GL(\mathbb{C}^2)$, and $\rho'_e$ denotes the derivative of the
representation map $\rho$ evaluated at the identity element
$e\in\newew$.

To proceed, an $ad$-invariant positive-definite inner product on
$u_{EW}(2)$ is required. In fact, there is a $2$-dimensional real
vector space of positive-definite $ad$-invariant bilinear forms on
$u_{EW}(2)$ given by (\cite{D})
\begin{equation}\label{bilinear}
  -\langle\bs{t}_{\alpha},\bs{t}_{\beta}\rangle
  =2g_1^{-2}\mathrm{Tr}(\bs{t}_{\alpha}\bs{t}_{\beta})
  +(g_2^{-2}-g_1^{-2})\mathrm{Tr}\bs{t}_{\alpha}
  \cdot\mathrm{Tr}\bs{t}_{\beta}
\end{equation}
for $\bs{t}_{\alpha},\bs{t}_{\beta}\in u_{EW}(2)$ where $g_1$ and
$g_2$ are real parameters.\footnote{That this bilinear form is
negative definite follows from
$2g_1^{-2}\mathrm{Tr}(\bs{t}_{\alpha}\bs{t}_{\beta})
  +(g_2^{-2}-g_1^{-2})\mathrm{Tr}\bs{t}_{\alpha}
  \cdot\mathrm{Tr}\bs{t}_{\beta}\leq 2g_1^{-2}\mathrm{Tr}(\bs{t}_{\alpha}\bs{t}_{\beta})
  +(g_2^{-2}-g_1^{-2})\mathrm{Tr}(\bs{t}_{\alpha}\bs{t}_{\beta})
  =(g_1^{-2}+g_2^{-2})\mathrm{Tr}(\bs{t}_{\alpha}\bs{t}_{\beta})< 0$.} A positive definite inner
product is obtained by the choice
\begin{equation}
  g_{\alpha\beta}:=(\bs{t}_{\alpha},\bs{t}_{\beta})
  \equiv -\langle\bs{t}_{\alpha},\bs{t}_{\beta}\rangle\;.
\end{equation}
Explicitly, in the basis defined by (\ref{basis}),
\begin{equation}
  g_{\alpha\beta}=
  \begin{pmatrix}
  0 & & g_W^{-2} & & 0 & & 0 \\
  g_W^{-2} & & 0 & & 0 & & 0 \\
  0 & & 0 & & g_Z^{-2} & & 0 \\
  0 & & 0 & & 0 & & g_Q^{-2} \\
  \end{pmatrix}
\end{equation}
where
\begin{equation}\label{inner product}
  g_W^{-2}:=(\bs{e}_{\pm},\bs{e}_{\mp}),\;
  g_Z^{-2}:=(\bs{h}_{1},\bs{h}_{1}),\;
  g_Q^{-2}:=(\bs{h}_{2},\bs{h}_{2})\;.
\end{equation}
The inner product can be put into canonical form by rescaling the
$u_{EW}(2)$ basis vectors by $\bs{e}_{\pm}\rightarrow
g_W\bs{e}_{\pm}$, $\bs{h}_{1}\rightarrow g_Z\bs{h}_{1}$, and
$\bs{h}_{2}\rightarrow g_Q\bs{h}_{2}$.

Since the Lie algebra is a direct sum of semisimple and abelian
algebras, the inner product defined on each Lie subalgebra is
proportional to the inner product for any of its faithful
representations. Hence, (\ref{matter fields}), (\ref{bilinear}),
(\ref{inner product}), and the orthogonality condition $u_{EM}\cap
\mathrm{k}=0$ give the $u_{EW}(2)$ representation (superscript +)
and the conjugate representation (superscript -) for the doublet
iquark matter fields;
\begin{eqnarray}\label{representation}
\bs{T}^{+}_{0}&=&\frac{ie}{2\cos{\Gt_W}\sin{\Gt_W}}\begin{pmatrix}
2\sin^{2}{\Gt_W}-1 & 0 \\ 0 & 1 \end{pmatrix},\notag \\
\bs{T}^{-}_{0}&=&\frac{-ie}{2\cos{\Gt_W}\sin{\Gt_W}}\begin{pmatrix}
1 & 0 \\ 0 & 2\sin^{2}{\Gt_W}-1 \end{pmatrix}, \notag \\
\bs{T}^{\pm}_{\pm}&=&\frac{ie}{\sqrt{2}\sin{\Gt_W}}\begin{pmatrix}
0 & 1 \\ 0 & 0
\end{pmatrix},\notag \\
\bs{T}^{\pm}_{\mp}&=&\frac{ie}{\sqrt{2}\sin{\Gt_W}}\begin{pmatrix}
0 & 0 \\ 1 & 0 \end{pmatrix},\notag \\
\bs{Q}^{+}&=&ie\begin{pmatrix} 1 & 0 \\ 0 &
0\end{pmatrix},\notag \\
\bs{Q}^{-}&=&-ie\begin{pmatrix} 0 & 0 \\ 0 & 1\end{pmatrix},
\end{eqnarray}
where $e$ is the electric charge, $\Gt_W$ is the Weinberg angle,
\begin{eqnarray}
  &g_Q^2&=\frac{g_1^2g_2^2}{(g_1^2+g_2^2)}=:e^2\nonumber\\
  &g_W^2&=\frac{g_1^2}{2}=:\frac{e^2}{2\sin^2\theta_W}\nonumber\\
  &g_Z^2&=\frac{(g_1^2+g_2^2)}{4}=\frac{e^2}{4\sin^2\theta_W\cos^2\theta_W}
\end{eqnarray}
and $r(s)$, $t$, and $v(u)$ were absorbed into $g_Q$,$g_W$, and
$g_Z$. This is, not surprisingly, identical to the SM left-handed
lepton representation.

The $1$-dimensional representation for the right-handed iquarks is
obtained by taking the trace of the $2$-dimensional representation
and using (\ref{basis}). For $h^{\pm}$ it amounts to taking the
trace of eq. (\ref{representation}). For the electrically neutral
$\xi^0$, it yields the trivial representation.

According to the discussion in Section \ref{charge and coupling
strength}, the inequivalent irreps of the direct product group
include the combinations $(\mathbf3,\mathbf2)$,
$(\mathbf3,\overline{\mathbf2})$, $(\mathbf3,\mathbf1^+)$,
$(\mathbf3,\mathbf1^-)$, and $(\mathbf3,\mathbf1^0)$ along with
the corresponding anti-particle combinations. We postulate that
the iquark matter field part of the Lagrangian density is
comprised of a sum over these combinations with appropriate
weights. The iquark contribution to the Lagrangian density is
therefore
\begin{eqnarray} \label{lagrangian}
\mathcal{L}_{\text{iquark}}&=&i\sum_{s}
 {\kappa^+(\overline{\Bold{H}}^+_{{\mathrm{L}},s}\D^+
 \Bold{H}^+_{{\mathrm{L}},s}+\overline{\Bold{H}}^+_{{\mathrm{R}},s}\D^+
 \Bold{H}^+_{{\mathrm{R}},s})}\nonumber\\
 &&\hspace{.4in}
 +\kappa^-(\overline{\Bold{H}}^-_{{\mathrm{L}},s}\D^-
 \Bold{H}^-_{{\mathrm{L}},s}+\overline{\Bold{H}}^-_{{\mathrm{R}},s}\D^-
 \Bold{H}^-_{{\mathrm{R}},s})
 +h.c.
\\ \nonumber\\
\mathcal{L}_{\text{Yukawa}}&=&-\sum_{s,t} m_{st}
\overline{\Bold{H}}^+_{{\mathrm{L}},s}
\mathit{\GP}^-\Bold{h}^+_{{\mathrm{R}},t} + n_{st}
\overline{\Bold{H}}^+_{{\mathrm{L}},s}
\mathit{\GP}^+\Bold{\xi}^0_{{\mathrm{R}},t}+h.c.
\end{eqnarray}
where $s,t$ label iquark generation. The matrices $m_{st}$ and
$n_{st}$ are general, generation(flavor) mixing mass matrices, and
$\mathit{\GP}^+$ is Higgs field
\begin{equation}
  \mathit{\GP}^+:=\begin{pmatrix}
  \phi^+ \\ \phi^0
  \end{pmatrix}\;.
\end{equation}

The covariant derivatives are
\begin{eqnarray} \label{covariant derivative}
\D^+ \Bold{H}^{+}_{\mathrm{L}} &=& \left(\not\!{\partial}+
\not\!\!{W}^{+}\bs{T}^{+}_{+}+ \not\!\!{W}^{-}\bs{T}^{+}_{-}+
\not\!\!{Z}^{0}\bs{T}^{+}_{0}
+ \not\!\!{A}\bs{Q}^{+}+\not\!G\bs{\Lambda}\right)\Bold{H}^{+}_{\mathrm{L}}\notag \\\nonumber\\
\D^- \Bold{H}^{-}_{\mathrm{L}} &=& \left(\not\!{\partial}+
\not\!\!{{W}^{+}}^*\bs{T}^{-}_{+}+
\not\!\!{{W}^{-}}^*\bs{T}^{-}_{-}+ \not\!\!{Z}^{0}\bs{T}^{-}_{0}
+ \not\!\!{A}\bs{Q}^{-}+\not\!G\bs{\Lambda}\right)\Bold{H}^-_{\mathrm{L}}\nonumber\\
\nonumber\\
 \D^{\pm} \Bold{h}^{\pm}_{\mathrm{R}}
&:=&
\mathrm{tr}\mathit[\D^{\pm}]\Bold{h}^{\pm}_{\mathrm{R}},\nonumber\\ \nonumber\\
\D\Bold{\xi}^0_{\mathrm{R}} &=&
\bigl(\not\!\partial+i\not\!{G}\bs{\Lambda}\bigr)\xi^0_{\mathrm{R}}\;,
\end{eqnarray}
where the trace is only over $\newew$ indices, and we have not
specified the representation $\bs{\Lambda}$ of $SU_C(3)$ since it
will not be needed.

The Yang-Mills, lepton and Higgs contributions to the Lagrangian
density are identical to the SM.

A few remarks are in order.

\begin{itemize}
\item  Re-scaling the iquark fields cannot
cancel the relative scale difference between $\Bold{H}^{+}$ and
$\Bold{H}^{-}$ since they are $\newew$ conjugate to each other
(unless $\kappa^+=\kappa^-$). Consequently, these factors are not
trivial and their ratio is not altered by renormalization. The
effect of the constants $\kappa^+$ and $\kappa^-$ is to re-scale
the charge $e$ in (\ref{representation}). Note that normalization
of the $SU_C(3)$ coupling strengths is not altered as long as
$\kappa^++\kappa^-=1$, That is, $\kappa^++\kappa^-=1$ guarantees
the $SU_C(3)$ intrinsic charge and coupling strength equality (see
\cite{LAC1}).

\item  The $\kappa^+$ and $\kappa^-$ terms in
$\mathcal{L}_{\mathrm{iquark}}$ are not invariant under distinct
$U(2)$; again because $\Bold{H}^{+}$ and $\Bold{H}^{-}$ are
$\newew$ conjugate to each other.

\item The $\xi_R^0$ fields completely decouple from the $\newew$
gauge bosons. However, they do couple to the $SU_C(3)$ gauge
bosons. They also have an induced mass due to the Higgs
interaction.

\end{itemize}

\subsection{Currents and Anomalies}
Using (\ref{matter fields}), (\ref{representation}) and
(\ref{lagrangian}), the $\newew$ currents for each iquark
generation work out to be\footnote{For clarity, the notation will
not indicate generation mixing. However, from now on, all iquark
fields are understood to be generation-mixed mass eigenstates; it
being understood that a concomitant Kobayashi-Maskawa type matrix
must now be included in $\mathcal{L}_{\mathrm{iquark}}$.}
\begin{eqnarray} \label{currents}
j^{0(Z)}_{\mu}&=&\frac{e}{2\sin{\Gt_W}\cos{\Gt_W}}
 \biggl[\kappa^+\Bigl(2\sin^{2}{\Gt_W}-1\Bigr)
\overline{h^{+}_{\mathrm{L}}}\Gg_{\mu}h^{+}_{\mathrm{L}}
\notag\\
 & & -\kappa^-\Bigl(2\sin^{2}{\Gt_W}-1\Bigr)
\overline{h^{-}_{\mathrm{L}}}\Gg_{\mu}h^{-}_{\mathrm{L}}\notag\\
 & & +\kappa^+2\sin^{2}{\Gt_W}\overline{h^{+}_{\mathrm{R}}}\Gg_{\mu}
h^{+}_{\mathrm{R}}+\kappa^+\overline{\Gx^{0}_{\mathrm{L}}}
\Gg_{\mu}\Gx^{0}_{\mathrm{L}}\notag\\
 & & -\kappa^-\overline{\Gx^{0}_{\mathrm{L}}}
\Gg_{\mu}\Gx^{0}_{\mathrm{L}}
-\kappa^-2\sin^{2}{\Gt_W}\overline{h^{-}_{\mathrm{R}}}
\Gg_{\mu}h^{-}_{\mathrm{R}}\biggr], \notag \\\notag \\
j^{0(A)}_{\mu}&=&
\kappa^+e\overline{h^{+}}\Gg_{\mu}h^{+}-\kappa^-e \overline{h^{-}}
\Gg_{\mu}h^{-},\notag\\\notag \\
j^{+}_{\mu}&=&\frac{e}{\sqrt{2}\sin{\Gt_W}} \biggl[
\kappa^+\overline{h^{+}_{\mathrm{L}}}\Gg_{\mu}\Gx^{0}_{\mathrm{L}}
+\kappa^-\overline{\Gx^{0}_{\mathrm{L}}}\Gg_{\mu}
h^{-}_{\mathrm{L}}\biggr],\notag \\\notag \\
j^{-}_{\mu}&=&\frac{e}{\sqrt{2}\sin{\Gt_W}} \biggl[
\kappa^+\overline{\Gx^{0}_{\mathrm{L}}}\Gg_{\mu}h^{+}_{\mathrm{L}}
+\kappa^-\overline{h^{-}_{\mathrm{L}}}\Gg_{\mu}
\Gx^{0}_{\mathrm{L}}\biggr]
\end{eqnarray}
where we used ${W^+}^*=W^-$ for the $\kappa^-$ terms in the two
charged currents and summation over color indices is implicit.

As is well known, for a consistent quantum version of this model
to exist, the anomalies associated with these currents must cancel
the lepton anomalies. Because the iquarks furnish the same electroweak
representation as the leptons and because there are three colors\footnote{By color, we mean $SU(3)$ degrees of freedom.} of each, one would not expect the anomalies in this model
to cancel trivially.

To check this, it is possible to use an isospin/hypercharge basis
in $\newew$. However, it is more direct to maintain the basis in
which the unbroken $U(1)$ is associated with electric charge. It
must be kept in mind that the $U_{EM}(1)$ quantities which enter
into the anomaly calculation are not the intrinsic electric
charges of the matter fields, per se, but the coupling strengths
in the photon/matter field current, $j^{0(A)}_\mu$. The
$U_{EM}(1)$, $SU_{W}(2)$ and $SU_C(3)$ contributions  of the
left-handed matter fields are given in Table \ref{table}.

\begin{table}[h]
\begin{tabular}{|l| c c c c c c c|}
\hline fermions &
\begin{math}(h^{+},\Gx^{0})_{\mathrm{L}}\end{math} &
\begin{math}(\Gx^{0},h^{-})_{\mathrm{L}}\end{math} &
\begin{math}\overline{h^{+}_\mathrm{R}}\end{math} &
\begin{math}\overline{h^{-}_\mathrm{R}}\end{math} &
\begin{math}\overline{\xi^{0}_\mathrm{R}}\end{math} &
\begin{math}(\nu^{0},l^{-})_{\mathrm{L}}\end{math} &
\begin{math}\overline{l^{-}_\mathrm{R}}\end{math} \\ \hline\hline
\begin{math}U(1)\end{math} & \begin{math}(\kappa^+,0)\end{math} &
\begin{math}(0,-\kappa^-)\end{math} &  \begin{math}-\kappa^+\end{math} &
\begin{math}\kappa^-\end{math} & \begin{math}0\end{math} &
\begin{math}(0,-1)
\end{math} & 1\\
\begin{math}SU(2)\end{math} & 2 & 2 & 1 & 1 & 1 & 2 & 1 \\
\begin{math}SU(3)\end{math} & 3 & 3 & \begin{math}\overline{3}
\end{math} & \begin{math}\overline{3}
\end{math} & \begin{math}\overline{3}\end{math} & 1 & 1 \\
\hline
\end{tabular}
\caption{Anomaly contributions for left-handed fermionic matter
fields.} \label{table}
\end{table}

There are only four cases to check including the gravitational
anomaly\cite{W3}: $[SU(2)]^2U(1)$, $[SU(3)]^2U(1)$, $[U(1)]^3$,
and $[G]^2U(1)$. In that order, the relevant terms are
\begin{subequations}
\begin{align}
\sum_{\text{doublets}} p&=3(\kappa^+)
+3(-\kappa^-)+(-1)=0\;,\label{non-trivial}\\
\sum_{\text{triplets}} p&=(\kappa^+) +(-\kappa^-)+(-\kappa^+)+
(\kappa^-)+0=0\;,\\
\sum_{\text{all}} p^{3}&=3(\kappa^+)^{3}
+3(-\kappa^-)^{3} +3(-\kappa^+)^{3} +3(\kappa^-)^{3}\notag\\
 & \ \ \ +3(0)^3+(-1)^{3}+(1)^{3}=0\;,\\ \notag \\
\sum_{\text{all}} p&=3(\kappa^+) +3(-\kappa^-)+3(-\kappa^+)
+3(\kappa^-)\notag\\ &  \ \ \ +3(0)+(-1)+(1)=0\;,
\end{align}
\end{subequations}
where $ep$ denotes the $U_{EM}(1)$ coupling parameter for the
iquark currents. With the exception of (\ref{non-trivial}), the
anomaly conditions are null rather trivially. From
(\ref{non-trivial}) and the condition $\kappa^++\kappa^-=1$, there
will be no anomaly associated with the gauge symmetries for the
choice
\begin{equation}\label{fractions}
\kappa^+=\frac{2}{3}\;,  \ \ \ \ \  \kappa^-=\frac{1}{3}\;.
\end{equation}

Now turn to the issue of the global chiral transformation,
\begin{eqnarray}
\Gd_{\Gl}h^{+}=i\Gl\Gg_{5}h^{+} &\ \ \ \ &
\Gd_{\Gl}\Gx^{0}=-i\Gl\Gg_{5}\Gx^{0}\;,
\end{eqnarray}
and the decay rate of $\pi^{0}\rightarrow 2\Gg$. The chiral
anomaly is proportional to
\begin{equation}
\mathrm{tr}\biggl[(\kappa^+ \bs{Q}^{+}+\kappa^-
\bs{Q}^{-})^{2}\bs{\tau}_{3}\biggr]
\end{equation}
which, for three colors, yields
\begin{equation}
3\Bigl(\frac{2}{3}\Bigr)^{2}-3\Bigl(\frac{1}{3} \Bigr)^{2}=1\;.
\end{equation}
Not surprisingly, this is identical to the result of the SM and
yields the correct decay rate.

To make contact with the SM, the conventional SM
generation-mixed quark mass eigenstates can be associated with
$h^{\pm}$ and $\Gx$. Inspection of (\ref{currents}) suggests that
the familiar fractionally charged quarks should be associated with
a pair of fields. Thus, make the following
correspondence:
\begin{subequations}\label{quarks}
\begin{eqnarray}
u^{+\frac{2}{3}}&\sim&{(h^{+},\Gx^{0})}\\
d^{-\frac{1}{3}}&\sim&{(\Gx^{0},h^{-})}
\end{eqnarray}
\end{subequations}
where $u$ and $d$ represent up and down quark fields respectively.
More accurately, the quark currents are identified with a pair of
$h^{\pm}$ and $\Gx^{0}$ currents:
\begin{eqnarray}
 \overline{u^{+\frac{2}{3}}_{L}}\Gg_{\mu}u^{+\frac{2}{3}}_{L}&\sim
&\left(\overline{h^{+}_{L}}\Gg_{\mu}h^{+}_{L}
 \,,\,\overline{\Gx^{0}_{L}}\Gg_{\mu}\Gx^{0}_{L}\right)
\nonumber\\
 \overline{d^{-\frac{1}{3}}_{L}}\Gg_{\mu}d^{-\frac{1}{3}}_{L}
 &\sim &\left(\overline{\Gx^{0}_{L}}\Gg_{\mu}\Gx^{0}_{L}\,,\,
\overline{h^{-}_{L}}\Gg_{\mu}h^{-}_{L}\right)
\nonumber\\
\overline{u^{+\frac{2}{3}}_{L}}\Gg_{\mu}d^{-\frac{1}{3}}_{L} &\sim
& \left(\overline{h^{+}_{L}}\Gg_{\mu}\Gx^{0}_{L}
\,,\,\overline{\Gx^{0}_{L}}\Gg_{\mu}h^{-}_{L}\right)\;.
\end{eqnarray}
So that, for example,
\begin{eqnarray}
\Bigl(\tfrac{4}{3}\sin^{2}{\Gt}-1\Bigr)
\overline{u^{+\frac{2}{3}}_{L}}\Gg_{\mu}u^{+\frac{2}{3}}_{L} \sim
\Bigl(\tfrac{4}{3}\sin^{2}{\Gt}&-&\tfrac{2}{3}\Bigr)
\overline{h^{+}_{L}}\Gg_{\mu}h^{+}_{L} -
\tfrac{1}{3}\overline{\Gx^{0}_{L}}\Gg_{\mu}\Gx^{0}_{L}\;,
\end{eqnarray}
and
\begin{eqnarray}
\overline{u^{+\frac{2}{3}}_{L}}\Gg_{\mu}d^{-\frac{1}{3}}_{L} \sim
 \tfrac{2}{3}\overline{h^{+}_{L}}\Gg_{\mu}\xi^0_{L} +
\tfrac{1}{3}\overline{\Gx^{0}_{L}}\Gg_{\mu}h^-_{L} \;.
\end{eqnarray}
There are analogous relations for the currents
$\overline{d^{-\frac{1}{3}}_{L}}\Gg_{\mu}d^{-\frac{1}{3}}_{L}$ and
$\overline{d^{-\frac{1}{3}}_{L}}\Gg_{\mu}u^{+\frac{2}{3}}_{L}$.
These correspondences will certainly lead to iquark current masses
and hadronic constituents that differ from the conventional quark
picture. Appendix A contains some \emph{tentative} iquark
composites for a selection of mesons and baryons.

To the extent that this correspondence is justified, the weak
currents in (\ref{currents}) agree with the SM currents.
Graphically, the correspondence associates a sum of one-particle
currents and their vertex factors with an equivalent two-particle
current, whose vertex factor is the sum of the individual vertex
factors of the constituent one-particle currents. Physically, the
correspondence constitutes an average description in the sense
that individual iquarks are not discriminated.

Although the electromagnetic current couples to the iquarks with
the correct fractional charge, it does not couple to all of the
energy-momentum carried by the iquarks since $\xi^0$ is
electrically neutral. However, as discussed later, it is expected
that $m_{h}/m_{\xi}\approx m_{l}/m_{\nu}$. So if this affects
transition probabilities, it would presumably be a small effect.

Evidently, the electroweak currents in (\ref{currents}) will agree
with the SM whenever: (i) individual quarks/iquarks cannot be
observed, and (ii) the experiment is not sensitive to the assumed
relatively small mass of $\Gx$. At this point in time, the first
case is ruled out. However, it is conceivable that some types of
experiments could be sensitive to the small mass ratio
$m_{\Gx}/m_{h}$. This might\footnote{To be sure, this model would
require different parton distribution functions than the SM, but
it is not clear if these would imply contradicting predictions for
scattering cross sections.} lead to disagreement with the SM that
would presumably become more prominent at high energy and large
momentum transfer.

\section{A Speculative Model}\label{speculative}
Even if it turns out that the iquark model cannot be
experimentally distinguished from the SM, it is theoretically
different because the leptons and iquarks furnish the same
$\newew$ representation. This fact begs the conjecture that
leptons and iquarks are just different manifestations of the same
underlying field.

Conventional wisdom is that $SU_C(3)\times U_{EM}(1)$ is unbroken
throughout the entire phase space. However, this assumes that
leptons and iquarks are separate matter fields.  It is conceivable
that matter fields exhibit different symmetry characteristics
dependent on space-time position, four-momentum, and particle
content. Perhaps leptons and iquarks are different phases of the
same matter field and that $SU_C(3)$ is broken in the lepton
phase.

Such a phase change would have to depend not only on the QCD
characteristic energy but also on the localized particle content.
Presumably then, the $SU(3)$ symmetry would require not only
sufficient energy but also the necessary iquark particle content
sufficiently localized in space-time. This, together with
asymptotic freedom, would conspire to suppress strong interactions
in typical lepton-hadron collisions.

It is interesting to implement this idea in terms of an effective
Lagrangian density. In this
speculative model, the Yang-Mills and Higgs contributions to the
Lagrangian density will be identical to the SM so they will not be displayed. Also, the Yukawa term is taken to be the
same as (\ref{lagrangian}) except there will be obvious
adjustments to the mass matrices. However, in place of the usual
iquark and lepton contributions, there will be a single
contribution denoted $\mathcal{L}_\mathrm{f}$. The relevant term
is
\begin{eqnarray} \label{new lagrangian 1}
\mathcal{L}_{\text{f}}&=&i\sum_{s}
\overline{\bs{F}}_{s}\not\!\mathrm{D} \bs{F}_{s}+h.c.
\end{eqnarray}
where
\begin{equation}
  \bs{F}:=\bs{S}\begin{pmatrix}
 \Bold{F}^+ \\ 0\end{pmatrix}\;,\;\;\;\begin{pmatrix}
  \Bold{F}^+ \\ \Bold{F}^-\end{pmatrix}\in\C^2\oplus\C^2\;,
\end{equation}
\begin{equation}\label{S}
  \bs{S}=\begin{pmatrix}
  \alpha & & \beta \\
  -\beta & & \alpha
  \end{pmatrix}\;,\;\;\;\bs{S}\bs{S}^{\mathrm{T}}=\mathbf{1}\;,
  \end{equation}
\begin{equation}
  \not\!\mathrm{D}:=\begin{pmatrix}
  \D^+ & & 0 \\
  0 & & (i\tau_2)(\D^-)(i\tau_2)^\dag
  \end{pmatrix}\;.
\end{equation}
$\not\!\!\mathrm{D}$ acts in the usual way on
left/right-handed fields, and includes $SU_C(3)\times\newew$ gauge
fields.

The matter fields $\Bold{F}$ are sections of $S\otimes
V_{\text{f}}$ where $V_{\text{f}}$ is the Whitney sum bundle of
the vector bundles associated with the representations furnished
by the left/right-handed $\Bold{F}$. As in previous sections,
$\Bold{F}^+$ and $\Bold{F}^-$ share the same space-time dependence
and spinor and color indices. Since $\Bold{F}^+=(i\tau_2)\Bold{F}^-$, then (\ref{new lagrangian 1})
reduces to
\begin{equation}\label{new lagrangian 2}
\mathcal{L}_{\text{f}}=i\sum_{s}\alpha^2\overline{\Bold{F}^+}\D^+\Bold{F}^+
+\beta^2\overline{\Bold{F}^-}\D^-\Bold{F}^-+h.c.\;.
\end{equation}

If there is a phase transition --- either induced by terms already
present in (\ref{new lagrangian 1}) or in an added term --- for some
regions of phase space, then the matter field phase space
$\mathcal{P}$ will have the form $\mathcal{P}=
\mathcal{P}_{SU_C(3)\times U_{EM}(1)}\cup\mathcal{P}_{U_{EM}(1)}$
with $\mathcal{P}_{SU_C(3)\times
U_{EM}(1)}\cap\mathcal{P}_{U_{EM}(1)}=\emptyset$ where
$\mathcal{P}_{SU_C(3)\times U_{EM}(1)}$ and
$\mathcal{P}_{U_{EM}(1)}$ represent field and canonical conjugate
field configurations of unbroken and broken $SU_C(3)$
respectively. In consequence, a functional integral over
$\mathcal{P}$ breaks into a sum of functional integrals over
$\mathcal{P}_{SU_C(3)\times U_{EM}(1)}$ and
$\mathcal{P}_{U_{EM}(1)}$.

Since the gauge field associated with $U_{EM}(1)$ exists
(presumably) continuously throughout $\mathcal{P}$, the conditions
of anomaly cancellation must hold everywhere in phase space. In
other words, photon exchange is possible for all charged matter
fields so $U_{EM}(1)$ currents must match across boundaries of
$\mathcal{P}_{SU_C(3)\times U_{EM}(1)}$ and
$\mathcal{P}_{U_{EM}(1)}$. Therefore, as a phase change occurs,
anomaly cancellation requires that $\bs{S}$ changes
discontinuously from $\alpha=1$, $\beta=0$ in
$\mathcal{P}_{U_{EM}(1)}$ to $\alpha=\sqrt{1/3}$,
$\beta=\sqrt{2/3}$ in $\mathcal{P}_{SU_C(3)\times U_{EM}(1)}$.

The theory can be conveniently reformulated in terms of a
functional integral over the full phase space $\mathcal{P}$ with
full $SU_C(3)\times\newew$ symmetry by introducing separate iquark
$(h^{\pm},\xi^0)$ and lepton $(l^{\pm},\nu^0)$ fields having
compact support on $\mathcal{P}_{SU_C(3)\times U_{EM}(1)}$ and
$\mathcal{P}_{U_{EM}(1)}$ respectively. In this case,
$\mathcal{L}_{\mathrm{f}}$ reduces to an effective
$\mathcal{L}_{\mathrm{iquark}}+\mathcal{L}_{\mathrm{lepton}}$ with
their associated $\alpha$ and $\beta$ values.

In any case, for a state evolving from a region
$\mathcal{P}_{SU_C(3)\times
U_{EM}(1)}\leftrightarrow\mathcal{P}_{U_{EM}(1)}$, it follows that
\begin{equation}\label{phase change}
  \begin{pmatrix}
  h^{+} \\ \xi^0
  \end{pmatrix}_s
  \longleftrightarrow
  \begin{pmatrix}
  l^{+} \\ \nu^0
  \end{pmatrix}_s
\end{equation}
implying: (i) a massive neutrino whose right-handed component
completely decouples except for gravity, (ii) the equivalence of
lepton and baryon electric charge, (iii) equal numbers of lepton
and iquark generations, (iv) $m_{h_s}/m_{\xi_s}\approx
m_{l_s}/m_{\nu_s}$ (ignoring renormalization effects), and (v) a
relationship between neutrino oscillations and iquark flavor
mixing. Additionally, it reduces the number of free parameters and
renders quark confinement a non-issue (or rather transmutes it
into a deconfinement issue).

The Yukawa term would imply that the bare lepton and iquark masses
are identical. However, renormalization due to self-energy
contributions destroys the degeneracy, and a (very) rough estimate
using $\alpha_s(m_Z)/\alpha_e(m_Z)\approx 10^1$ suggests
$m_{h_s}\approx 10m_{l_s}$.

There is of course one glaring drawback to this idea: Where are the $SU(3)$ Goldstone bosons in the lepton phase? One way to wriggle out of this issue is to assume the iquark/lepton phase change does not break $SU(3)$.

\section{Summary}

We presented an alternative representation for quarks in the
Standard Model. Central to the motivation is the idea that, within
a given representation, elementary particle and antiparticle
states with multiple charges associated with local internal
symmetries should realize all possible charge combinations. This
leads to the possibility of non-trivial factors multiplying
certain matter field terms in the Lagrangian density.

Implementing this idea within the context of the Standard Model
leads to iquarks with integer electric charge that, nevertheless,
couple to the photon with fractional charges. The resulting model
differs from the Standard Model, because some of the iquarks (with
small mass) do not couple to the photon. However, it is not clear
if the difference is experimentally detectible.

The fact that the iquarks and leptons furnish the same $\newew$
representation suggests that they are manifestations of the same
underlying field. This would seem to require a phase change in
certain regions of field phase space. If this characterization
turns out to be correct, it would relate some of the parameters of
the SM, and, hopefully, aid in the search for an underlying
theory.

\appendix
\section{Mesons and Baryons} The mesons and baryons will be
composites of $\Bold{H}_s\overline{\Bold{H}}_t$ and
$\Bold{H}_s\Bold{H}_t\Bold{H}_u$ respectively where $s,t,u$ label
iquark generation. Although each component of
$\Bold{H}:=(\Bold{H}^+,\Bold{H}^-)$ is a spinor, we will omit the
various combinations of spin for simplicity. Since
$\Bold{H}^{\pm}$ has two (electroweak) components, a little work
is required to exhibit the elementary field content of the
composites.

Table \ref{mesons} contains \emph{tentative} pseudoscalar
assignments for the first two generations of composites
$\Bold{H}_1^{\pm}\overline{\Bold{H}^{\pm}}_1$,
$\Bold{H}_1^{\pm}\overline{\Bold{H}^{\pm}}_2$, and
$\Bold{H}_2^{\pm}\overline{\Bold{H}^{\pm}}_2$. Denote the top
component of $\Bold{H}$ by $\wedge$ and the bottom component by
$\vee$. Meson composites should correspond to combinations of
$\ket{\wedge\overline{\wedge}}$, $\ket{\vee\overline{\vee}}$, and
$\tfrac{1}{\sqrt{2}}\ket{\wedge\overline{\vee}+\vee\overline{\wedge}}$,
(ignoring spin combinations). There, of course, will be many more
possible combinations and admixtures of spin and generation than
those represented in the table.

Table \ref{baryons} contains assignments of selected spin $1/2$
and $3/2$ baryons to iquark composites
$\Bold{H}_1^{\pm}\Bold{H}_1^{\pm}\Bold{H}_1^{\pm}$,
$\Bold{H}_1^{\pm}\Bold{H}_1^{\pm}\Bold{H}_2^{\pm}$,
$\Bold{H}_1^{\pm}\Bold{H}_2^{\pm}\Bold{H}_2^{\pm}$, and
$\Bold{H}_2^{\pm}\Bold{H}_2^{\pm}\Bold{H}_2^{\pm}$. For
simplicity, the table displays only the iquark doublet content.

It should be emphasized that the composite assignments in the
tables are only tentative.


\begin{table}[p]
\begin{tabular}{|c c c c l l|}
\hline & Meson &&   doublet composite&  & iquark composition
 \\ \hline
& $\pi^0$ &&  $H_1^{\pm}\overline{H_1^{\pm}}$& & $\sim(h_1^
+\overline{\xi^0_1}+\xi_1^0\overline{h_1^+})+$ C.
\\
& $\pi^+$ &&  $H_1^+\overline{H_1^-}$& &
$\sim(h_1^+\overline{\xi_1^0}+\xi_1^0\overline{h^-_1})$
\\
& $\pi^-$ &&  $H_1^-\overline{H_1^+}$& &
$\sim(h_1^-\overline{\xi_1^0}+\xi_1^0\overline{h^+_1})$
\\&&&&&\\
& $\left.\begin{array}{c}K^0\\D^0\end{array}\right\}$ &&
$H_1^{\pm}\overline{H_2^{\pm}}+1\leftrightarrow 2$& &
$\sim\left\{\begin{array}{c}
(h_1^+\overline{\xi^0_2}+\xi_2^0\overline{h_1^+})+\mathrm{C}.\\(h_2^
+\overline{\xi^0_1}+\xi_1^0\overline{h_2^+})+\mathrm{C}.\end{array}\right.$
\\
& $\left.\begin{array}{c}K^+\\D^+\end{array}\right\}$ &&
$H_1^+\overline{H_2^-}+1\leftrightarrow 2$& &
$\sim\left\{\begin{array}{c}
(h_1^+\overline{\xi_2^0}+\xi_2^0\overline{h^-_1})\\
(h_2^+\overline{\xi_1^0}
+\xi_1^0\overline{h^-_2})\end{array}\right.$
\\
& $\left.\begin{array}{c}K^-\\D^-\end{array}\right\}$ &&
$H_1^-\overline{H_2^+}+1\leftrightarrow 2$& &
$\sim\left\{\begin{array}{c}
(h_1^-\overline{\xi_2^0}+\xi_2^0\overline{h^+_1})\\
(h_2^-\overline{\xi_1^0}
+\xi_1^0\overline{h^+_2})\end{array}\right.$
\\&&&&&\\
& ${D_s^*}$ &&  $H_2^{\pm}\overline{H_2^{\pm}}$& & $\sim(h_2^
+\overline{\xi^0_2}+\xi_2^0\overline{h_2^+})+$ C.
\\
& $D_s^+$ &&  $H_2^+\overline{H_2^-}$& &
$\sim(h_2^+\overline{\xi_2^0}+\xi_2^0\overline{h^-_2})$
\\
& $D_s^-$ &&  $H_2^-\overline{H_2^+}$& &
$\sim(h_2^-\overline{\xi_2^0}+\xi_2^0\overline{h^+_2})$
\\
\hline
\end{tabular}
\caption{Iquark assignments for selected mesons. For the iquarks,
the overline denotes an antifield, $\pm$ and $0$ denote electric
charge, the subscript denotes the iquark generation, and C. stands
for $\newew$ conjugate. Summation over color indices is implied.}
\label{mesons}
\end{table}

\begin{table}[p]
\begin{tabular}{|c c c c c|}
\hline & Baryon &  &  & doublet composite
 \\ \hline
& $p$ &  & & $\Bold{H}_1^+\Bold{H}_1^+\Bold{H}_1^-$
\\
& $n$ &  & & $\Bold{H}_1^+\Bold{H}_1^-\Bold{H}_1^-$
\\
& $\Delta^-$ &  & & $\Bold{H}_1^-\Bold{H}_1^-\Bold{H}_1^-$
\\
& $\Delta^{++}$ &  & & $\Bold{H}_1^+\Bold{H}_1^+\Bold{H}_1^+$
\\
& $\Sigma^+$ &  & & $\Bold{H}_1^+\Bold{H}_1^+\Bold{H}_2^-$
\\
& $\Sigma^0,\Lambda$ &  & & $\Bold{H}_1^+\Bold{H}_1^-\Bold{H}_2^-$
\\
& $\Sigma^-$ &  & & $\Bold{H}_1^-\Bold{H}_1^-\Bold{H}_2^-$
\\
& $\Xi^0$ &  & & $\Bold{H}_1^+\Bold{H}_2^-\Bold{H}_2^-$
\\
& $\Xi^-$ &  & & $\Bold{H}_1^-\Bold{H}_2^-\Bold{H}_2^-$
\\
& $\Sigma_c^{++}$ &  & & $\Bold{H}_1^+\Bold{H}_1^+\Bold{H}_2^+$
\\
& $\Sigma_c^+,\Lambda_c^+$ &  & &
$\Bold{H}_1^+\Bold{H}_1^-\Bold{H}_2^+$
\\
& $\Sigma_c^0$ &  & & $\Bold{H}_1^-\Bold{H}_1^-\Bold{H}_2^+$
\\
& $\Xi_c^+$ &  & & $\Bold{H}_1^+\Bold{H}_2^-\Bold{H}_2^+$
\\
& $\Xi_c^0$ &  & & $\Bold{H}_1^-\Bold{H}_2^-\Bold{H}_2^+$
\\
& $\Omega_c^0$ &  & & $\Bold{H}_2^-\Bold{H}_2^-\Bold{H}_2^+$
\\
& $\Xi_{cc}^{++}$ &  & & $\Bold{H}_1^+\Bold{H}_2^+\Bold{H}_2^+$
\\
& $\Xi_{cc}^+$ &  & & $\Bold{H}_1^-\Bold{H}_2^+\Bold{H}_2^+$
\\
& $\Omega^-$ &  & & $\Bold{H}_2^-\Bold{H}_2^-\Bold{H}_2^-$
\\
& $\Omega_{cc}^+$ &  & & $\Bold{H}_2^-\Bold{H}_2^+\Bold{H}_2^+$
\\
& $\Omega_{ccc}^{++}$ &  & &
$\Bold{H}_2^+\Bold{H}_2^+\Bold{H}_2^+$
\\
\hline
\end{tabular}
\caption{Iquark doublet assignments for selected spin $1/2$ and
$3/2$ baryons.} \label{baryons}
\end{table}

\end{document}